\documentclass[prl,reprint,superscriptaddress,longbibliography,floatfix,twocolumn, nofootinbib]{revtex4-2}
\usepackage[utf8]{inputenc}
\usepackage{amsmath,graphicx, dsfont, bm}
\usepackage[normalem]{ulem}
\usepackage{diagbox}
\usepackage{fancyhdr, amssymb}
\usepackage[T1]{fontenc}

\usepackage{xcolor}

\newcommand{\ket}[1]{\left| {#1} \right\rangle}
\newcommand{\bra}[1]{\left\langle {#1} \right|}
\newcommand{\braket}[2]{\langle #1 | #2 \rangle }

\newcommand{\prjct}[1]{\mathinner{|{#1}\rangle}\!\!\mathinner{\langle{#1}|}}
\newcommand{\tr}{\text{tr}\,}

\newcommand{\id}{\mathds{1}}

\newcommand{\cE}{\mathcal{E}}
\newcommand{\cF}{\mathcal{F}}
\newcommand{\cH}{\mathcal{H}}
\newcommand{\cM}{\mathcal{M}}

\renewcommand{\t}[1]{\mathrm{#1}}
\newcommand{\be}{\begin{equation}}
\newcommand{\ee}{\end{equation}}

\newcommand{\Fi}{R}

\usepackage{nicefrac}

\begin{document}

\title{
Calibration-Independent Certification of a Quantum Frequency Converter
}

\author{Matthias Bock}
\affiliation{Experimentalphysik, Universität des Saarlandes, 66123 Saarbrücken, Germany}
\affiliation{Institut für Experimentalphysik, Universität Innsbruck, Technikerstra{\ss}e 25, 6020 Innsbruck, Austria}
\author{Pavel Sekatski}
\affiliation{Department of Applied Physics, University of Geneva, 1205 Geneva, Switzerland}
\author{Jean-Daniel Bancal}
\affiliation{Institut de Physique Théorique, Université Paris-Saclay, CEA, CNRS, 91191 Gif-sur-Yvette, France}
\author{Stephan Kucera}
\affiliation{Experimentalphysik, Universität des Saarlandes, 66123 Saarbrücken, Germany}
\author{Tobias Bauer}
\affiliation{Experimentalphysik, Universität des Saarlandes, 66123 Saarbrücken, Germany}
\author{Nicolas Sangouard}
\affiliation{Institut de Physique Théorique, Université Paris-Saclay, CEA, CNRS, 91191 Gif-sur-Yvette, France}
\author{Christoph Becher}
\affiliation{Experimentalphysik, Universität des Saarlandes, 66123 Saarbrücken, Germany}
\author{J{\"u}rgen Eschner}
\affiliation{Experimentalphysik, Universität des Saarlandes, 66123 Saarbrücken, Germany}

\date{\today}

\begin{abstract}
We report on a method to certify a unitary operation with the help of source and measurement apparatuses whose calibration throughout the certification process needs not be trusted. As in the device-independent paradigm our certification method relies on a Bell test, but it removes the need for high detection efficiencies by including the single additional assumption that non-detected events are independent of the measurement settings. The relevance of the proposed method is demonstrated experimentally with the certification of a quantum frequency converter. The experiment starts with the heralded creation of a maximally entangled two-qubit state between a single $^{40}$Ca$^+$ ion and a 854 nm photon. Entanglement preserving frequency conversion to the telecom band is then realized with a non-linear waveguide embedded in a Sagnac interferometer. The resulting ion-telecom photon entangled state is characterized by means of a Bell-CHSH test from which the quality of the frequency conversion is quantified. We demonstrate the successful frequency conversion with an average certified fidelity of $\geq 84\%$ and an efficiency $\geq 3.1\times 10^{-6}$ at a confidence level of 99\%. This ensures the suitability of the converter for integration in quantum networks from a trustful characterization procedure.
\end{abstract}

\maketitle


\paragraph{Introduction--}
The enabling technologies for the realization of networks capable of linking quantum systems together have been identified~\cite{Kimble2008, Sangouard2011, Wehner2018}. This includes quantum frequency converters -- nonlinear processes in which a photon of one frequency is converted to another frequency whilst preserving all other quantum properties. A converter acts as a quantum photonic adapter allowing one for example to interface high-energy photonic transitions of quantum matters with lower-energy photons better suited for long-distance travel. Together with quantum storage and processing devices, quantum frequency converters enable a range of new technologies using quantum networks, from distributed quantum computing~\cite{Jiang2007}, quantum-safe cryptography~\cite{Pirandola2020}, enhanced sensing~\cite{Proctor2018, Sekatski2020} and time keeping~\cite{Komar2014}.

\medskip

A natural question arising in view of this integration potential is how to certify the functioning of a quantum frequency converter independently of contingent details, i.e.~without the need to know an exhaustive physical model of its inner functioning or to assume that the certification equipment (source and measurements) is well calibrated and remains perfectly calibrated for the whole duration of the certification procedure. Recent works have demonstrated that the quantum nature of a number of channels can be witnessed with assumptions on the source calibration but without any trust on the measurement apparatus~\cite{Pusey:15, PhysRevX.8.021033, YongYu,Graffiti}.  However, it would be desirable to quantify the quality of the device. Indeed, an ideal certification method should ensure the usability of the converter for all future purposes. A radical solution to this task is offered by the method of device-independent (DI) characterization, also known as self-testing~\cite{Supic2020}, where the physical implementation of a device is inferred from the correlations observed in a Bell-type experiment~\cite{Brunner2014}. The device-independent approach relies on the separation and independence between the apparatuses at hand, but makes no assumption on their internal modeling. As far as we know, only two self-tests have been fully implemented experimentally to date, both related to state certification~\cite{Tan2017, Bancal2021}. The main reason for this scarcity is that device-independent certification is very demanding regarding the efficiency of measurement apparatuses~\cite{Vertesi2010}. This requirement has been circumvented in a number of experimental state certifications based on post-selected Bell inequality violation -- by considering only the statistics observed from detected events~\cite{Wang2018, Gomez2019, Goh2019}. The question of what remains device-independent in certifications using post-selections has not been discussed in these experimental realizations.

\medskip
In this article, we provide an accessible method to certify trustworthily unitary operations. 
Inspired by the device-independent certification techniques presented in Refs.~\cite{magniez, blocks}, our method certifies the quality of a unitary operation from a single Bell test but without requiring high detection efficiencies to be implemented. Precisely, we assume that the physical process responsible for the occurrence of no-detection events is independent of the choice of the measurement setting, but impose no further restriction on it~ \cite{Orsucci2020howpostselection}. Therefore, no-detection events may still depend on the state being measured or on devices' calibrations in an arbitrary way. This natural assumption allows us to substantially reduce the complexity of unitary certification by removing the need for high overall detection efficiencies without requiring to trust the calibration of the certification devices. We use this tool to realize the first calibration-independent certification of a unitary -- a state-of-the-art polarization-preserving quantum frequency converter (QFC)~\cite{Leent2022, Arens2023, Krut2019, Ikuta2018}. We employ a trapped-ion platform as source of light-matter entanglement between an atomic Zeeman qubit and the polarization state of a spontaneously emitted photon~\cite{Kurz2016}. A frequency conversion based on a highly-efficient difference frequency generation process in a nonlinear waveguide embedded in a polarization Sagnac interferometer connects the system wavelength at 854\,nm to the telecom C-band at 1550\,nm~\cite{Arens2023}. A Bell-CHSH test~\cite{CHSH1969} is finally performed after the frequency conversion, using the ion-telecom photon entangled state. We demonstrate the successful frequency conversion with an average certified fidelity of $\geq$ 84\% and a probability to get a telecom photon detection conditioned on a successful ion state readout of $3.1 \times 10^{-6}$ at a confidence level of 99\%.


\bigskip

\paragraph{Source, QFC and measurement apparatus modelling--}
We start by providing an "a priori" quantum model of several devices involved in the setup. The desired models rely on minimal assumptions on the internal functioning of the devices, which nevertheless have enough physical insight to describe the process of frequency conversion.

\medskip

A QFC can be represented by a channel between two physical systems identified as its input and output. In our case, these are the photonic modes entering the QFC device from the source and exiting it towards the detectors.  These modes are filtered to ensure that their frequencies lie in the desired bandwidth $\omega_i$ and $\omega_f$ respectively. We can associate to these photonic modes  two Hilbert spaces $\cH_A^{(i)}$ and $\cH_A^{(f)}$, that encompass all the degrees of freedom necessary to describe the emission of the source in addition to the frequencies. To describe the quality of the QFC to be characterized, we are only interested in how the device maps the photonic states received from the source to the state it sends to the detector:
\be\label{eq: real QFC}
\text{QFC}: B(\cH_A^{(i)}) \to B(\cH_A^{(f)}).
\ee
Here $B(\cH_A^{(i)})$ stands for the set of bounded operators on the Hilbert space $\cH_A^{(i)}$ (similarly for $B(\cH_A^{(f)})$).
All auxiliary outputs can be safely ignored, while auxiliary input systems are to be seen as part of the device\footnote{Note for example that the QFC is powered by a laser stimulating the difference frequency generation process, which in turn requires energy supply. These are required for the proper functioning of the device, and have to be seen as parts of the QFC.}. The completely positive and trace-preserving (CPTP) map QFC in Eq.~\eqref{eq: real QFC} is unknown and our goal is to provide a recipe to characterize it.

\medskip

In addition to the QFC itself, our setup involves an entanglement source preparing a state shared between two parties called Alice (A) and Bob (B). Alice's system is carried by the electromagnetic field used to characterize the QFC, which is associated to Hilbert space $\cH_A^{(i)}$ introduced above. The physics of Bob's system is irrelevant for the purpose of the QFC characterization because the QFC resides entirely on Alice's side. Its state spans a Hilbert space $\cH_B$. We denote the state produced by the source by
\be
\rho^{(i)} \in B(\cH_A^{(i)}\otimes \cH_B).
\ee
The state obtained after applying the converter on Alice's side reads 
\be
\rho^{(f)} = \left(\text{QFC}\otimes \text{id}\right) [\rho^{(i)}] \in B(\cH_A^{(f)}\otimes \cH_B).
\ee

\medskip

Finally, the form of the quantum model for the measurement apparatus is needed in order to describe the occurrences of the measurement results. We introduce two possible measurements  $\cM_A^{(f)}$ and $\cM_B$  which act on the system of Alice after the converter and the system of Bob, respectively. Ideally, the measurements should have binary inputs $x,y = 0,1$ and binary outputs $a,b = 0,1$. In practice however, a third outcome  $a,b =\emptyset$ is possible corresponding to a no-click event. Each of the measurements is given by two POVMs with three elements each, such as
\be
\cM_A^{(f)} \simeq \{M_{a|x}^{(f)}\}, \quad \cM_B \simeq \{M_{b|y}\}
\ee
with the operators $M_{a|x}^{(f)}$ and $M_{b|y}$ acting on $\cH^{(f)}_A$ and $\cH_B$ respectively.

\bigskip

\paragraph{Weak fair-sampling assumptions--} Following the results presented in Ref.~\cite{Orsucci2020howpostselection}, we now introduce an assumption on the measurement structure which allows us to relax the requirement on the detection efficiency inherent to device-independent certification. Consider for example the measurement $\cM_B$ specified by the POVM elements $M_{b|y}$ with settings $y$ and outcomes $b$ including a no-click outcome $b=\emptyset$. 
The measurement $\cM_B$ satisfies the \textit{weak fair-sampling assumption} if
\begin{equation}\label{eq:wfs}
M_{\emptyset|y} = M_{\emptyset|y'},
\end{equation}
i.e.~the occurrence of the no-click outcome is not influenced by the choice of the measurement setting\footnote{Note that the weak fair-sampling assumption is much less restrictive than the usual (strong) fair-sampling assumption and hence much easier to enforce in practice. Indeed, in addition to the weak fair-sampling condition, the strong fair-sampling assumption further requires that all POVM elements associated to no-click events should be a multiple of the identity operators. This imposes a strong structure on the underlying measurement, implying in particular that its behavior must be independent of the measured state, which is not the case for the weak fair-sampling assumption.}. Under this assumption, $\cM_B$ can be decomposed as a filter $\Fi_B$ acting on the quantum input (a quantum instrument composed of a completely positive (CP) map and a failure branch that outputs $y=\emptyset$) followed by a measurement $\overline{\cM_B}$ with unit efficiency (without  $b=\emptyset$ output)~\cite{Orsucci2020howpostselection}, that is 
\be
\cM_B = \overline{\cM_B} \circ \Fi_B.
\ee

\medskip

Assuming that Bob's measurement fulfills the weak fair-sampling assumption, we can focus on the data post-selected on Bob's successful detection only. This data can be associated to an experiment where a probabilistic source prepares a state 
\be
\varrho^{(i)}=\frac{(\t{id}\otimes \Fi_B)[\rho^{(i)}]}{\tr (\t{id}\otimes \Fi_B)[\rho^{(i)}]},
\ee
conditional on the successful outcome of Bob's filter $\Fi_B$. We can therefore only consider the experimental runs where the state $\varrho^{(i)}$ is prepared and Bob's detector clicks.

\medskip

Assuming that Alice's measurement also fulfills the weak fair-sampling assumption, that is 
\be
\cM_A^{(f)}= \overline{\cM}_A^{(f)}\circ \Fi_A,
\ee
we perform a similar decomposition for the final state 
\be\label{eq: rhof}
\varrho^{(f)} = \frac{\left((\Fi_A\!\circ\!\text{QFC})\otimes \t{id}\right)[\varrho^{(i)}]}{\tr \left((\Fi_A\!\circ\!\text{QFC})\otimes \t{id}\right)[\varrho^{(i)}]},
\ee
corresponding to post-selected events for both sides. The success rate $\text{P}_\text{succ}(\Fi_A) = \tr ((\Fi_A\circ\text{QFC})\otimes \t{id})[\varrho^{(i)}]$ of the filtering $\Fi_A$ is given by the probability to observe a click event on Alice's detector, conditional on the click event seen by Bob (defining $\varrho^{(i)}$)
\be\label{eq: Psucc FA}
\text{P}_\text{succ}(\Fi_A)= \text{P}(\text{click at Alice}|\text{click at Bob}).
\ee

\bigskip

\paragraph{Goal--} To set our goal, we first specify what a frequency converter is expected to do. While changing the carrier frequency from $\omega_i$ to $\omega_f$, an ideal QFC should not affect any other degree of freedom carrying meaningful information. Therefore, in the case of photons encoding a qubit degree of freedom within their polarization,  an ideal QFC should act as the identity
\be\label{eq: ideal QFC}
\t{id}_2: B(\mathds{C}^2) \to B(\mathds{C}^2)
\ee 
on the polarization of a single photon. We thus need  to show that, while changing the frequency of the photons, the map \eqref{eq: real QFC} is capable of preserving a two-dimensional subspace.

\medskip

Following Ref.~\cite{blocks}, this can be formalized by requiring the existence of two maps $V: B(\mathds{C}^2) \to B(\cH_A^i)$ (injection map) and $ \Lambda:  B(\cH_{A}^i) \to B(\mathds{C}^2)$ (extraction map) such that 
\be\label{eq: certificate}
\Lambda \circ\t{QFC}\circ V \approx \t{id}_2,
\ee
where the approximate sign refers to a bound on the Choi fidelity between the two maps $\cF(\cE,\cE')= F\left((\text{id}\otimes\cE)[\Phi^+], (\text{id}\otimes\cE')[\Phi^+]\right)$, where $\Phi^+$ is a maximally entangled two-qubit state  and $F(\rho,\sigma)= \left(\tr|\sqrt{\rho} \sqrt{\sigma}|\right)^2$ is the fidelity between two states $\rho$ and $\sigma$. Note that in the case where $\cE'$ is the identity map, Choi fidelity takes a particularly simple form $\cF(\cE,\text{id}) = \bra{\Phi^+} (\text{id}\otimes\cE)[\Phi^+] \ket{\Phi^+}$.
\medskip

We are concerned with non-deterministic frequency converters. More precisely, our goal is thus to compare the actual frequency converter to a probabilistic but heralded quantum frequency converter -- a device which behaves as an ideal QFC with a certain probability, and otherwise reports a failure. To do so, we can allow the  maps $\Lambda$ and $V$ to be non trace preserving. The quality of an announced QFC is captured by two parameters --  the probability that it works and the error it introduces in this case. These are quantified by the following figures of merit. The success probability 
\be\label{eq: sucess}
P_\text{succ}(\Lambda \circ \text{QFC}\circ V) = \tr \left((\Lambda \circ \text{QFC}\circ V)\otimes \text{id}\right) [\Phi^+],
\ee
captures the efficiency of the converter. The conditional Choi fidelity
\be\label{eq: fidelity}
\cF(\Lambda \circ \text{QFC}\circ V)= \bra{\Phi^+}\frac{\left((\Lambda \circ \text{QFC}\circ V)\otimes \text{id}\right) [\Phi^+]}{P_\text{succ}(\Lambda \circ \text{QFC}\circ V) }\ket{\Phi^+},
\ee
bounds the error introduced in the state conditional to a successful frequency conversion. Certifying the converter thus consists in establishing lower-bounds on both quantities $P_\text{succ}$ and $\cF$.

\bigskip

\paragraph{Certification--} Following Ref.~\cite{blocks}, we certify the QFC through the self-testing of the maximally entangled two qubit state $\Phi^+$ derived in Ref.~\cite{jed}. The latter is based on the Clauser Horne Shimony Holt (CHSH) inequality -- a well-known Bell test derived for a setting where two parties Alice and Bob can choose one of two binary measurements at each round. The CHSH score $S$ is given by
\be\label{eq: CHSH first}
S = \sum_{a,b,x,z=0,1}(-1)^{a+b+xy}P(a,b|x,y),
\ee
where $a,b=0,1$ are the parties' measurement outcomes, and $x,y=0,1$ label their measurement setting. In the quantum framework, the correlation $P(a,b|x,y)$ is given by $P(a,b|x,y)=\tr \rho\,  M_{a|x}^A\otimes M_{b|y}^B$ where $\rho$ is the measured state and $\{M_{a|x}^A\}$ ($\{M_{b|y}^B\}$) are Alice (Bob)'s appropriate POVM elements.
We know from Ref.~\cite{jed} that for any quantum model $(\rho, M_{a|x}^A, M_{b|y}^B)$ exhibiting a CHSH score $S$, there exist local extraction maps $\Lambda_A$ and $\Lambda_B$ such that $\bra{\Phi^+}(\Lambda_A\otimes \Lambda_B)[\rho] \ket{\Phi^+}\geq f(S)$ for
\be\label{eq: f(S)}
f(S) = \frac{12 + (4+5\sqrt{2})(5S-8)}{80}.
\ee
Notably, the form of the maps $\Lambda_{A(B)}$ does not depend on the measurement performed by the other party.

\medskip

This result holds for all quantum states and measurement. When applying it to the quantum model of the filtered state $\varrho^{(f)}$ after the QFC and the binary measurements $\overline{\cM}_A^{(f)}$ and $\overline{\cM}_B$ for instance, it implies that there exist local maps $\overline{\Lambda}_A^{(f)}$ and $\overline{\Lambda}_B$ such that
\be
\label{eq: Ffinal}
\bra{\Phi^+}(\overline{\Lambda}_A^{(f)} \otimes \overline{\Lambda}_B)[\varrho^{(f)}] \ket{\Phi^+} \geq f(S),
\ee
where $S$ is the CHSH score of the binary measurements on the filtered state, i.e. the post-selected CHSH score.

\medskip

To derive a certificate on the QFC itself rather than of its output state, we need to show that the state before the action of the QFC can be prepared from $\Phi^+$ with the injection map $V_A$ acting on Alice, i.e.~$(\text{id}\otimes \overline{\Lambda_B})[\varrho^{(i)}] \approx (\id\otimes V_A) [\Phi^+]$. We show in the Methods that this can be done perfectly, i.e.
\be\label{eq: injection}
(\text{id}\otimes \overline{\Lambda_B})[\varrho^{(i)}] = \frac{ (V_A \otimes \text{id}) [\Phi^+]}{\tr (V_A \otimes \text{id}) [\Phi^+]},
\ee
with a probabilistic map $V_A$ associated to the success rate $\text{P}_\text{succ}(V_A) = \tr (V_A \otimes \text{id}) [\Phi^+] \geq 50\%$. This is possible because the state $(\text{id}\otimes \overline{\Lambda_B})[\varrho^{(i)}]$ is carried by a qubit at Bob's side. It can therefore be purified to a state of Schmidt rank 2 and any such state can be efficiently obtained from $\Phi^+$ by a local filter applied by Alice.

\medskip

Combining the definition of the filtered state $\varrho^{(f)}$ in Eq.~\eqref{eq: rhof} with Eqs.~\eqref{eq: Ffinal} and \eqref{eq: injection}, we conclude that for the probabilistic extraction map $\Lambda_A= \overline{\Lambda}_A^{(f)}\circ \Fi_A$, the conditional Choi fidelity of Eq.~\eqref{eq: fidelity} is bounded by
\be
\label{eq:choiState}
\cF(\Lambda_A \circ \text{QFC}\circ V_A)\geq f(S).
\ee
We emphasize that this bound is valid for all possible underlying state $\rho$ and measurements $\{M_{a|x}^A\}$, $\{M_{b|y}^B\}$ subject to Eq.~\eqref{eq:wfs}.

\medskip

It remains to bound the success probability of the map $\Lambda_A \circ \text{QFC}\circ V_A$ when applied on $\Phi^+$, that is $\text{P}_\text{succ}(\Lambda_A \circ \text{QFC}\circ V_A)= \tr \left((\overline{\Lambda}_A^{(f)}\circ \Fi_A\circ \text{QFC}\circ V_A)\otimes \text{id}\right)[\Phi^+]$. This map is successful if both the injection map $V_A$ and the filter $\Fi_A$ are, hence 
\be\begin{split}\label{eq:Psucc}
\text{P}_\text{succ}(\Lambda_A \circ \text{QFC}\circ V_A) &= \text{P}_\text{succ}(\Fi_A) \text{P}_\text{succ}(V_A) \\ &\geq \frac{1}{2} \text{P}_\text{succ}(\Fi_A). 
\end{split}
\ee
$\text{P}_\text{succ}(\Fi_A)$ can be estimated experimentally using Eq.~\eqref{eq: Psucc FA}.

\bigskip

\paragraph{Experimental source of entanglement--}

\begin{figure*}[tb]
    \centering
    {\includegraphics[width=0.93\linewidth]{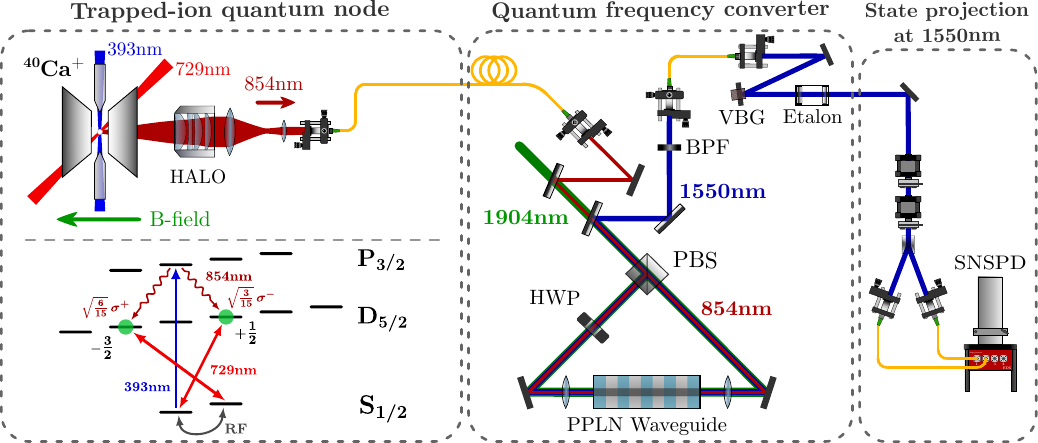}} 
     \caption{(a) Experimental setup. Light-matter entanglement is generated between a single trapped $^{40}$Ca$^{+}$ ion and the polarization state of an emitted photon at 854\,nm. The photons are collected with a high-aperture laser objective (HALO), coupled to a single-mode fiber and guided to the QFC device. The latter features a PPLN waveguide embedded in a polarization Sagnac interferometer to guarantee polarization-preserving operation. The converted photons pass a series of spectral filters (band-pass filter (BPF), volume Bragg grating (VBG) and etalon) to suppress background stemming from the DFG process. The projection setup at 1550\,nm consists of a motorized QWP and HWP, a Wollaston prism to split orthogonally-polarized photons, and two fiber-coupled superconducting-nanowire single-photon detectors (SNSPD). In the lower left part the level scheme of the $^{40}$Ca$^+$ ion including the most relevant states and transitions for entanglement generation and quantum state readout is shown. The atomic qubit is encoded in two Zeeman levels ($m = -3/2$ and $m = 1/2$) of the metastable $D_{5/2}$-state.}\label{fig:ExpSetup}
\end{figure*}

The experimental setup is sketched in Fig. \ref{fig:ExpSetup}.
Our source of entanglement is a trapped-ion quantum network node which creates light-matter entanglement between a Zeeman qubit in a single trapped $^{40}$Ca$^{+}$ ion (Bob) and the polarization state of an emitted single photon at 854\,nm (Alice) \cite{Bock2018IPE}. The photons are coupled to a single-mode fiber via a high-aperture laser objective (HALO) and guided to the frequency converter, which is the device we aim to certify.

\medskip

The entanglement generation sequence is slightly modifed compared to \cite{Bock2018IPE}. The relevant level scheme for the state preparation and detection of the Ca ion is shown in Fig \ref{fig:ExpSetup}. After Doppler cooling, excitation of the ion on the $S_{1/2}$ to $P_{3/2}$ transition by a $\pi$-polarized, 2\,$\mu$s long laser pulse at 393\,nm creates a spontaneously emitted photon at 854\,nm. This photon is collected along the quantization axis, thereby suppressing $\pi$-polarized photons, and is entangled with the ion in the state
\begin{equation}     
\ket{\Psi} = \sqrt{\frac{2}{3}}\,\ket{\sigma^{+}, \downarrow} + \sqrt{\frac{1}{3}}\,e^{i\omega_L\, t}\,\ket{\sigma^{-}, \uparrow}
\end{equation}     
with $\ket{\downarrow} = \ket{D_{5/2}, m = -\nicefrac{3}{2}}$ and $\ket{\uparrow} = \ket{D_{5/2}, m = +\nicefrac{1}{2}}$. The oscillation with frequency $\omega_L$ arises from the frequency difference between the $\ket{\uparrow}$ and $\ket{\downarrow}$ states and the asymmetry in the state results from the different Clebsch-Gordan coefficients (CGC) of the transitions between the $\ket{P_{3/2}}$ and $\ket{D_{5/2}}$ Zeeman sublevels. We compensate for this by means of a partial readout of the trapped-ion Zeeman qubit during the state preparation: a $\pi/2$-pulse at 729\,nm transfers 50\,\% of the population from $\ket{\downarrow} = \ket{D_{5/2}, m = -\nicefrac{3}{2}}$ to the $S_{1/2}$ ground state. 
A subsequent fluorescence detection with the cooling lasers is a projective measurement of this population in the following way. The fluorescence detection discriminates between population in the $S_{1/2}$-state which results in scattering of photons from the cooling laser, while population in $D_{5/2}$ leaves the ion dark. If it yields a bright result, the measurement is discarded, while a dark result leaves the D-state intact and heralds a successful state preparation. Thus, the ion-photon state after a dark result is maximally entangled
\begin{equation}\label{eq: atom photon state}     
\ket{\Psi} = \sqrt{\frac{1}{2}}\,\left(\ket{\sigma^{+}, \downarrow} + e^{i\omega_L\, t}\,\ket{\sigma^{-}, \uparrow}\right) \ .
\end{equation}     
In this way, maximally-entangled ion-photon pairs are generated at a rate of 720\,$s^{-1}$ and a probability per shot of 0.36\,\%. 

\bigskip

\paragraph{Experimental QFC device--} The QFC device transduces the photons at 854\,nm to the telecom C-band at 1550\,nm via the difference frequency generation (DFG) process 1/854\,nm - 1/1904\,nm = 1/1550\,nm in a periodically-poled lithium niobate (PPLN) waveguide~\cite{Arens2023}. The input photons are overlapped with the classical pump field at 1904\,nm on a dichroic mirror and guided to the core of the QFC device, an intrinsically phase-stable polarization Sagnac interferometer. The latter ensures polarization-preserving operation since the DFG process is inherently polarization-selective. The interferometer is constructed in a similar way as in \cite{Leent2020APE}, i.e.~a polarizing beam-splitter (PBS) spatially separates the orthogonal components and a HWP rotates the not convertible horizontal component of input, pump and output fields by 90$^{\circ}$. Both components are subsequently coupled to the same waveguide from opposite directions. The converted photons take the same interferometer paths, are recombined in the PBS, separated from the pump and input photons via another dichroic mirror and coupled to a single-mode fiber. Multi-stage spectral filtering down to 250\,MHz suppresses pump-induced background photons stemming from anti-Stokes Raman scattering in the waveguide. The external device efficiency is measured to 57.2\,\%, independent of the polarization and including all losses between input and output fiber.
The QFC-induced background is measured at the operating point to be 24(3)\,photons/s, being to our knowledge the lowest observed background of a QFC device in this high-efficiency region. 

\bigskip

\paragraph{Measurements--} To measure the Bell parameter $S$, we perform joint measurements of the atomic and photonic qubit in the four CHSH basis settings which we choose to lie in the equatorial plane of the Bloch sphere with respect to the basis defined in Eq.~\eqref{eq: atom photon state}.

\medskip

For the atomic qubit, the required basis rotation is implemented by means of a pulsed sequence of two consecutive $\pi$-pulses at 729\,nm and a radio-frequency (RF) $\pi/2$-pulse applied on the $S_{1/2}$ ground-state qubit with phase $\phi_{RF}$ using a resonant magnetic field coil (Fig. \ref{fig:ExpSetup}). The ground-state qubit states are readout by means of two fluorescence detection rounds yielding bright and dark events depending on whether the state is populated or not, respectively. The phase of the atomic qubit underlies the Larmor precession in $D_{5/2}$ with $\omega_L$. The arrival time $t$ of the photon reveals this Larmor phase up to a constant offset which is calibrated with an independent measurement and kept fixed for all following ones (see next section).

\medskip

For the photonic qubit we employ a set of a motorized quarter- and a half-wave plate for arbitrary basis rotations and a Wollaston prism to split orthogonally polarized photons. Both outputs are connected to fiber-coupled superconducting-nanowire singe-photon detectors (SNSPDs). To fulfill the weak fair sampling assumption we have to balance the efficiencies of both detectors since the error of the post-selected probabilities scales linearly with the imbalance. To this end we use attenuated laser light and adjust the bias current through the SNSPDs to achieve $\gamma = 1 - \eta_{snspd 1}/\eta_{snspd 2} \leq 0.2\,\%$ with $\eta_{snspd 2} = 13.5\,\%$. This reduces the deviation of the post-selected probabilities from those obtained with a lossless detector to about 1\,\%~\cite{Orsucci2020howpostselection}.

\medskip

To avoid influences of drifts over the measurements, we consecutively acquire runs of data for 5 seconds in each basis and cascade up to 660 runs. The CHSH score is then obtained using the setting choices 
\be
x=0 \rightarrow \frac{1}{2}(\sigma_x+1), \quad 
x=1 \rightarrow \frac{1}{2}(\sigma_y+1) 
\ee
in the photonic side and 
\be
y=0 \rightarrow \frac{1}{2}(\frac{\sigma_{x} + \sigma_{y}}{\sqrt{2}}+1), \quad 
y=1 \rightarrow \frac{1}{2}(\frac{\sigma_{x} - \sigma_{y}}{\sqrt{2}}+1)
\ee
for the ion side.

\bigskip

\begin{figure}
\centering
{\includegraphics[width=\linewidth]{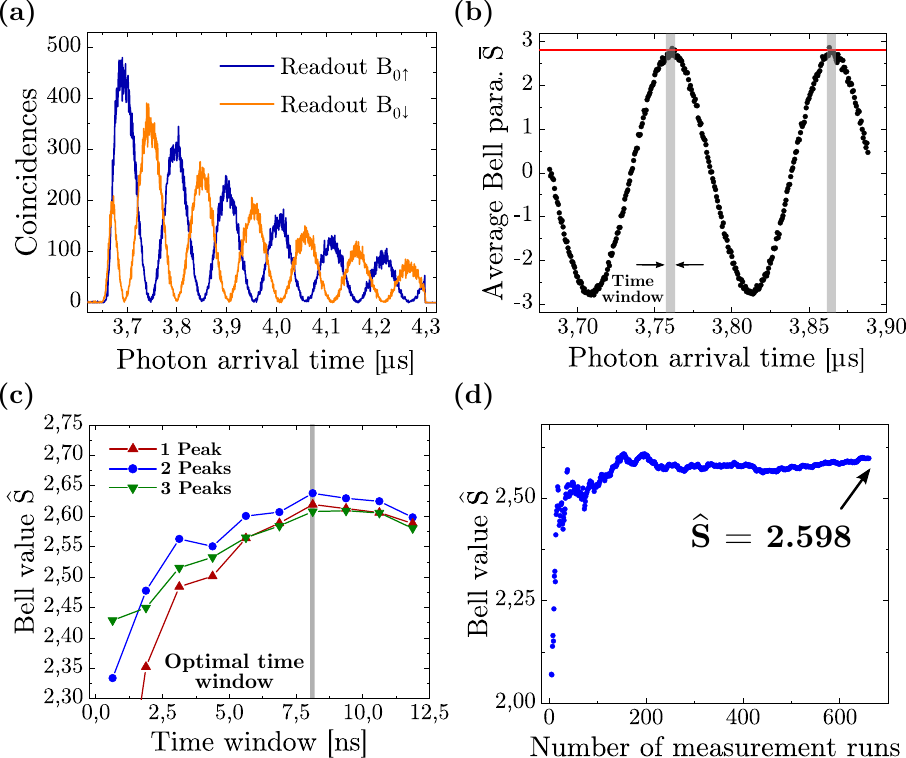}} 
\caption{(a) Time-resolved coincidences between photonic detection events of one of the detectors (readout base A$_0$) and bright events of the atomic state readout from both orthogonal states of the readout base B$_0$. The oscillations with the Larmor frequency of the atomic qubit stem from the time-dependency of the entangled state. (b) The average Bell parameter in dependence of the detection time. For the further analysis we select coincidences at the detection times which correspond to a Larmor phase of $\omega_L\, t = \pi$ resulting in the $\Psi^{-}$ Bell state. To obtain a feasible SBR, we only select concidences in time windows around the first two maxima, whose positions were determined from an independent measurement. (c) The optimal Bell value is a tradeoff between number of detected events (favouring large time windows) and phase resolution (favouring small time windows). (d) Bell value after QFC after a certain number of measurement runs based on the coincidences from the optimal time window.}\label{fig:resst}
\end{figure}

\paragraph{Experimental results--} Figure \ref{fig:resst}(a) shows a typical time-resolved coincidence histogram between photonic detection events of one of the detectors (readout base $x=0$) and bright events of the atomic state readout (base $y=0$). As mentioned previously, the oscillations stem from the Larmor precession of the atomic qubit resulting in a time-dependent entangled state, Eq. \ref{eq: atom photon state}. From the histograms of all readout bases we calculate the CHSH Bell parameter according to Eq.~\eqref{eq: CHSH first}, which is consequently also detection-time dependent (see figure \ref{fig:resst}(b)). Thus, by postselecting coincidences in a certain time window, we perform a readout in the correct CHSH basis. These windows are located at the top of each oscillation, they are calibrated from an independent measurement and kept fixed during the analysis.

\medskip

To certify the QFC from finite experimental data, we view the multi-round experiment as a sequence of rounds, labelled with $i=1,\ldots,n$. Note that each round is an experimental trial of atom-photon-state generation, not the previously mentioned runs.
At each round $i$, the final atom-photon state corresponds to some intrinsic CHSH score $S_i$. By virtue of Eqs.~\eqref{eq:choiState} the conditional Choi fidelity of the converter at round $i$ satisfies $\cF_i \geq f(S_i)$, with $f$ given in Eq.\eqref{eq: f(S)}.
We are interested to bound the average fidelity
\begin{equation}
    \overline \cF = \frac{1}{n}\sum_{i=1}^n \cF_i
\end{equation}
over all measurement rounds.
By linearity of $f$, this quantity is bounded by $f(\overline S)$, where $\overline S=\frac{1}{n}\sum_{i=1}^n S_i$ is the average CHSH score. A lower bound on $\overline S$ thus lower-bounds $\overline \cF$ through Eq.~\eqref{eq: f(S)}.

\medskip

To give a clear lower bound on the CHSH score $\overline S$ in presence of a finite number of measurement rounds, we construct a one-sided confidence interval on $\overline S$. It can be shown that
\begin{equation}
    \hat S = 8 I_\alpha^{-1}(n\overline T, n(1-\overline T) + 1) - 4
\end{equation}
is the tightest such lower bound for a confidence level $1-\alpha$ whenever $\alpha<1/4$ and $n,n\overline T\geq 2$~\cite{inPreparation}. Note that this conclusion does not rely on the I.I.D. assumption (independent and identically distributed), e.g.~it holds true even if the state produced by the setup is not identical at each round. Here, $\overline T=\sum_{i=1}^n T_i$ is the experimental mean of the random variables corresponding to the CHSH game
\begin{equation}
    T_i = \begin{cases}
    1 & A_i\oplus B_i = X_i Y_i\\
    0 & A_i\oplus B_i \neq X_i Y_i\\
    \end{cases}.
\end{equation}
where $X_i/A_i$ ($Y_i/B_i$) is Alice's (Bob's) measurement setting/outcome in round $i$. Setting $\alpha=0.01$, we obtain 99\%-confidence lower bounds $\hat S$ on $\overline S$ for the state produced in the experiment.

\medskip

Fig. \ref{fig:resst}(c) show the calculated Bell values $\hat S$ -- obtained from the independent calibration measurement -- for different numbers of time windows (located at each oscillation peak) and different window lengths. The optimal values are a tradeoff between a higher number of events favouring better statistics and thus higher Bell values, the signal-to-background ratio which decreases with an increasing number of peaks due to the exponential decay of the photon wavepacket, and phase resolution being ideal for the smallest possible time window. We choose an optimal time window of 8.125\,ns (corresponding to 9 time bins) and the first and second oscillation peak.

\medskip

The final results of the certification are displayed in Fig. \ref{fig:resst}(d): we see the Bell value $\hat S$ after each measurement run (in each run we measure all four correlators). We find a converging behavior due to the increasing number of events, which reduces the statistical uncertainty. The remaining fluctuations of the Bell values after QFC are most likely caused by drifts of the unitary rotation of the photon polarization state in the fiber connecting the ion trap and QFC setup. After 660 runs (16593 events) we find 
\begin{equation}
   \hat S = 2.598 
\end{equation}
and an average observed CHSH score of $8 \overline T-4 =2.65$. The latter is in good agreement with our known error sources, namely signal-to-background-ratio (0.04), phase resolution (0.028), atomic coherence and fidelity of the atomic state readout (0.085), and polarization drifts in the long fiber (0.028). From $\hat S$, we calculate via Eq.~\eqref{eq: f(S)} and Eq.~\eqref{eq:choiState} a certified conversion fidelity of
\begin{equation}
   \hat \cF \geq f(\hat S)  = 0.8406   
\end{equation}

To conclude, we bound the converter's efficiency. Eq.~\eqref{eq:Psucc} and Eq.~\eqref{eq: Psucc FA} allow us to bound the success probability of the QFC directly as a function of the number of coincidence detection at Alice and Bob $n_c$ and the total number of rounds $n$ as
\begin{equation}
    \hat P_\text{succ} = \frac{1}{2} I^{-1}_\alpha(n_c, n-n_c+1),
\end{equation}
where we used the probability estimator free of the I.I.D. assumption from Ref.~\cite{inPreparation}. With $n_c=16\,593$ and $n=2\,640\,000\,000$, we obtain the lower bound
\begin{equation}
    P_\text{succ} \geq \hat P_\text{succ} = 3.1 \times 10^{-6}
\end{equation}
at a confidence level $1-\alpha=99\%$. The limited overall success probability can be attributed to several factors: the success probability to collect a photon at 854\,nm from the ion (0.36\,\%), the external device efficiency of the QFC, i.e.~the probability to get a 1550\,nm photon in the output fiber per 854\,nm photon at the input of the QFC (57\,\%), the quantum efficiency of the single-photon detectors (13.5\,\%), further optical losses in the whole experimental setup (60\,\%) and the ratio between the post-selected time window and the total photon wavepacket (3.9\,\%).

\vspace{0.2em}

\bigskip

\paragraph{Conclusion--} We have presented the first recipe leveraging device-independent techniques to certify a unitary operation without assuming that the certification devices are perfectly calibrated. Although not fully device-independent, the proposed recipe is widely tolerant to loss. This is achieved by assuming that the occurrence of no-detection events is independent from the choice of measurement, which is both more general and more realistic than independence from the measured state. We used the calibration-independent method on a state-of-the-art polarization-preserving quantum frequency converter to demonstrate its performance in term of conversion efficiency and fidelity. The proposed recipe could be used to certify quantum storage and processing devices among others. Given the interesting balance between its practical feasibility and high level of trust, we believe that the method is well suited to become a reference certification technique to ensure the suitability of devices for their integration in quantum networks.

\bigskip

\paragraph{Note added--} While writing this manuscript we became aware of a related experimental work by Neves et al.~\cite{Neves23}.

\bigskip

\paragraph{Acknowledgement--} J-D.B. and N.S. acknowledge support by the Institut de Physique Th\'eorique (IPhT), Commissariat à l’Energie Atomique et aux Energies Alternatives (CEA), and by the European High-Performance Computing Joint Undertaking (JU) under grant agreement No 101018180 and project name HPCQS. J.E., S.K., C.B., M.B. and T.B. acknowledge support by the German Federal Ministry of Education and Research (BMBF) through projects Q.Link.X (16KIS0864), and QR.X (16KISQ001K). Furthermore, M.B. acknowledges funding from the European Union’s Horizon 2020 research and innovation programme under the Marie Sklodowska-Curie grant agreement No 801110 and the Austrian Federal Ministry of Education, Science and Research (BMBWF).

\paragraph{Author Contributions--} M.B. and S.K. conceived the experiments. P.S., J.-D.B. and N.S. developed the theoretical concepts. M.B., P.S., J.-D.B. and S.K. analyzed the data. M.B., S.K. and T.B. contributed to the experimental setup. P.S., J.-D.B., M.B., S.K. and N.S. prepared the manuscript with input from all authors. N.S., C.B. and J.E. jointly supervised the project. 

\paragraph{Competing interests--} The authors declare no competing interests.

\section*{Methods}

In this section we present a detailed derivation of the Eqs.~(\ref{eq:choiState},\ref{eq:Psucc}) of the main text. We start by recalling the context in which they apply.

Consider a scenario where Alice and Bob control quantum systems, described by unknown finite dimensional Hilbert spaces $\cH_A^{(i)}$ and $\cH_B$, prepared in a global state  $\varrho^{(i)}$. In addition, Alice has access to a probabilistic channel modeled by some completely positive trace non increasing map $\Fi_A\circ \text{QFC}: B(\cH_A^{(i)})
\to B(\cH_A^{(f)})$. When successfully applied on Alice's system the channel outputs the state 
\be
\varrho^{(f)} = \frac{\left((\Fi_A\!\circ\!\text{QFC})\otimes \t{id}\right)[\varrho^{(i)}]}{\tr \left((\Fi_A\!\circ\!\text{QFC})\otimes \t{id}\right)[\varrho^{(i)}]},
\ee
which is self-tested to be close to a maximally entangled two qubit state $\ket{\Phi^+} = \frac{1}{\sqrt 2}(\ket{00}+\ket{11})$. That is, there exist completely positive trace preserving maps $\bar \Lambda_A^{(f)}: B(\cH_A^{(f)}) \to B(\mathds{C}^2)
)$ and $\bar \Lambda_B: B(\cH_B) \to B(\mathds{C}^2)
)$ such that
\be\label{eq: fid bound}
\bra{\Phi^+}  (\bar\Lambda_A^{(f)}\otimes  \bar \Lambda_B)[ \varrho^{(f)} ] \ket{\Phi^+} \geq f(S).
\ee
Let us now show that these predicates are sufficient to guarantee the results of Eqs.~(\ref{eq:choiState},\ref{eq:Psucc}) discussed in the main text.

\subsection{State preparation}

First, let us define the states
\begin{align}
\bar{\varrho}^{(i)} &= (\text{id}\otimes \bar \Lambda_B)[\varrho^{(i)}]\\
\label{eq: st f}
\bar \varrho^{(f)} &= \frac{\left((\Fi_A\!\circ\!\text{QFC})\otimes \t{id}\right)[\bar \varrho^{(i)}]}{\tr \left((\Fi_A\!\circ\!\text{QFC})\otimes \t{id}\right)[\bar \varrho^{(i)}]},
\end{align}
which are positive semi-definite operators on the Hilbert space $\cH_A^{(i)}\otimes \mathds{C}^2$ and $\cH_A^{(f)}\otimes \mathds{C}^2$ respectively. These definitions allow us to rewrite the Eq.~\eqref{eq: fid bound} in the form
\be\label{eq: st f 2}
\bra{\Phi^+}  (\bar\Lambda_A^{(f)}\otimes \text{id})[ \bar \varrho^{(f)} ] \ket{\Phi^+} \geq f(S),
\ee
with the map $\bar \Lambda_B$ absorbed in the state. To be able to interpret this bound as Choi fidelity let us now show that the initial state $\bar{\varrho}^{(i)}$ can be prepared from $\Phi^+$ with the help of a local probabilistic map applied by Alice.

To do so, introduce an auxilliary quantum system $A'$ and consider a purification of the state 
\be\label{eq: purif}
\bar{\varrho}^{(i)} = \t{tr}_{A'} \prjct{\Psi},
\ee
where $\prjct{\Psi}$ is a pure state on $\cH_A^{(i)} \otimes \cH_{A' } \otimes\mathds{C}^2$. Since Bob's system is a qubit by Schmidt theorem this state is of the form 
\be
\ket{\Psi} = \sum_{k=0}^{1} \lambda_k \ket{\xi_k}_{AA'} \ket{b_k}_B ,
\ee
with $\lambda_0\geq \lambda_1$ and orthogonal states $\braket{b_0}{b_1}=\braket{\xi_0}{\xi_1}$. There is thus a qubit unitary $u_B$ and an isometry $v_{AA' }: \mathds{C}^2 \to \cH_A^{(i)} \otimes \cH_{A' } $ such that 
\be
\ket{\Psi} = (v_{AA' }\otimes u_B)   \left(\lambda_0 \ket{00} + \lambda_1 \ket{11}\right).
\ee
In addition, it is straightforward to see that the following qubit filter (completely positive trace non increasing map) with
\be\begin{split}
\Fi'_A : B(\mathds{C}^2) &\to B(\mathds{C}^2)\\
\rho &\mapsto K \rho K\\
\text{with}\quad K = & \left(\begin{array}{cc}
1&\\
& \frac{\lambda_1}{\lambda_0}
\end{array}\right)
\end{split}
\ee
satisfies 
\be
\frac{(v_{AA' }\circ \Fi_A' \otimes{u_B})[\Phi^+]}{\text{tr}(v_{AA' }\circ \Fi'_A \otimes{u_B})[\Phi^+]} = \Psi
\ee and has success probability
\be
\text{tr}(v_{AA' }\circ \Fi' _A \otimes{u_B})[\Phi^+] = \frac{1}{2} + \frac{1}{2}\left(\frac{\lambda_1}{\lambda_0}\right)\geq \frac{1}{2}.
\ee
Combining with Eq.~\eqref{eq: purif} and using  $(\t{id}\otimes u_B)[\Phi^+] = (u_B^T\otimes \t{id})[\Phi^+]$we conclude that the probabilistic map
\be\begin{split}
V_A : B(\mathds{C}^2) &\to B(\cH_A^{(i)})\\
\rho &\mapsto \tr_{A'} (v_{AA' }\circ \Fi'_A \circ u_B^T)[\rho]
\end{split}\ee
fulfills 
\be\label{eq: state prep}
\frac{(V_A \otimes \t{id})[\Phi^+]}{\text{tr}(V_A \otimes \t{id})[\Phi^+]} = \bar \varrho^{(i)}
\ee
and has success probability at least 50\% when acting on $ \Phi^+$. Demonstrating the desired result.\\

\subsection{Certifying the QFC}

Combining Eqs.~(\ref{eq: st f},\ref{eq: st f 2},\ref{eq: state prep}) we obtain the following bound
\be
\bra{\Phi^+}  \frac{(\bar\Lambda_A^{(f)}\!\circ\!\Fi_A\!\circ\!\text{QFC}\!\circ\!  V_A \otimes \text{id})[ \Phi^+ ] }{\text{tr}(\bar\Lambda_A^{(f)}\!\circ\!\Fi_A\!\circ\!\text{QFC}\!\circ\!  V_A \otimes \text{id})[ \Phi^+ ]}\ket{\Phi^+} \geq f(S).
\ee
Which has the form of a bound on the Choi fidelity of the channel $\bar\Lambda_A^{(f)}\!\circ\!\Fi_A\!\circ\!\text{QFC}\!\circ\!  V_A$ with respect to the identity channel. To get the expression of the main text it we to define a single probabilistic extraction map 
$
\Lambda_A = \bar \Lambda_A^{(f)}\!\circ\!\Fi_A
$
in order to obtain
\be
\cF(\Lambda_A \circ \text{QFC}\circ V_A) \geq f(S).
\ee
It remains to argue about the minimal possible value of the success probability. By virtue of $(V_A \otimes \t{id})[\Phi^+] = \bar \varrho^{(i)} \, {\text{tr}(V_A \otimes \t{id})[\Phi^+]}$ we get 
\be
\begin{split}\nonumber
    &\text{P}_\text{succ}(\Lambda_A \circ \text{QFC}\circ V_A) \\
     &=\text{tr}(\bar\Lambda_A^{(f)}\!\circ\!\Fi_A\!\circ\!\text{QFC}\!\circ\!  V_A \otimes \text{id})[ \Phi^+ ] \\
    &= ( \tr (\bar\Lambda_A^{(f)}\!\circ\!\Fi_A\!\circ\!\text{QFC}  \otimes \text{id})[\bar \varrho^{(i)}])(\text{tr}(V_A \otimes \t{id})[\Phi^+]) \\
    &\geq \frac{1}{2} \tr (\bar\Lambda_A^{(f)}\!\circ\!\Fi_A\!\circ\!\text{QFC}  \otimes \text{id})[\bar \varrho^{(i)}]
    \\
    & = \frac{1}{2} \tr (\bar\Lambda_A^{(f)}\!\circ\!\Fi_A\!\circ\!\text{QFC}  \otimes \bar \Lambda_B)[ \varrho^{(i)}]\\
    &=  \frac{1}{2} \tr (\Fi_A\!\circ\!\text{QFC}  \otimes \text{id})[ \varrho^{(i)}]\\
    & = \frac{1}{2}\text{P}_\text{succ}(F_A) =  \frac{1}{2}\text{P}(\text{click at Alice}|\text{click at Bob})
\end{split}
\ee
where we used the fact that the maps $\bar\Lambda_A^{(f)}$ and $ \bar \Lambda_B$ are trace preserving. We thus conclude that the success probability of the map $(\Lambda_A \circ \text{QFC}\circ V_A)$ is at most twice lower than the conditional click probability of Alice $\text{P}(\text{click at Alice}|\text{click at Bob})$, which can be directly estimated from experimental data.

\end{document}